\documentclass[preprint,journal]{vgtc}

\usepackage{xspace}
\usepackage{listings}
\usepackage[svgnames]{xcolor}

\newcommand{\ie}{{i.e.,}\xspace}
\newcommand{\eg}{{e.g.,}\xspace}

\newcommand{\ea}{{et~al.}\xspace}

\newcommand{\etc}{{etc.}\xspace}

\newcommand{\bpstart}[1]{\vspace{1mm} \noindent{\textbf{#1.}}}

\newcommand{\bstartnc}[1]{\vspace{1mm} \noindent{\textbf{#1}}}

\definecolor{codeGreen}{cmyk}{0.85,0.2,1,0.15}
\definecolor{darkblue}{cmyk}{0.95,0,0,0.5}

\lstdefinelanguage{DRACO}{
  keywords={preference, helper, violation, entity, attribute},
  keywordstyle=\color{blue},
  identifierstyle=\color{codeGreen},
  sensitive=false,
  comment=[l]{\%},
  commentstyle=\color{darkgray}\ttfamily,
  stringstyle=\color{darkblue}\ttfamily,
  numberstyle=\color{darkblue}\ttfamily,
  morestring=[b]',
  morestring=[b]"
}

\usepackage{amsmath,amssymb,amsthm}

\DeclareMathOperator{\hugeE}{\mbox{\huge\raise-0.3ex\hbox{E}}}
\DeclareMathOperator{\p}{\mathbb{P}}
\DeclareMathOperator{\hugep}{\mbox{\huge\raise-0.3ex\hbox{$\p$}}}

\vgtccategory{Research}
\vgtcpapertype{algorithm/technique}

\title{Data Augmentation for Visualization Design Knowledge Bases}

\author{%
  \authororcid{Hyeok Kim}{0000-0003-4340-4470} and \authororcid{Jeffrey Heer}{0000-0002-6175-1655}
}

\authorfooter{
  \item
  	Hyeok Kim and Jeffrey Heer are with the University of Washington.
  	E-mail: \{hyeokk, jheer\}@uw.edu.
}

\abstract{%
    Visualization knowledge bases enable computational reasoning and recommendation over a visualization design space.
    These systems evaluate design trade-offs using numeric weights assigned to different features (\eg~binning a variable).
    Feature weights can be learned automatically by fitting a model to a collection of chart pairs, in which one chart is deemed preferable to the other.
    To date, labeled chart pairs have been drawn from published empirical research results; however, such pairs are not comprehensive, resulting in a training corpus that lacks many design variants and fails to systematically assess potential trade-offs.
    To improve knowledge base coverage and accuracy, we contribute data augmentation techniques for generating and labeling chart pairs.
    We present methods to generate novel chart pairs based on design permutations and by identifying under-assessed features---leading to an expanded corpus with thousands of new chart pairs, now in need of labels.
    Accordingly, we next compare varied methods to scale labeling efforts to annotate chart pairs, in order to learn updated feature weights.
    We evaluate our methods in the context of the Draco knowledge base, demonstrating improvements to both feature coverage and chart recommendation performance.
}


\teaser{
  \centering
  \includegraphics[width=\linewidth, alt={This figure has four sections, numbered from A to D.
Section A. Original design pairs can be obtained from empirical studies or design examples.
There is a stack of pairs of charts.
Each design pair consists of a more preferred design with a thumb-up icon and a less preferred design with a thumb-down icon. 
Section B. Primitive augmentation.
Enumerating design pairs showing similar differences in design primitives given an original design pair. 
There is a stack of chart pairs. 
From a chart from section A, a design difference about faceting by column vs. row is extracted. 
Those chart pairs are enumerated by altering the other primitives.
Section C. Feature augmentation.
Enumerating design pairs with design properties that the original pairs do not cover. 
From Section A, uncovered properties are derived.
They are having columns with X, and Binning a color channel.
Then, these properties are ablated to produce a chart pair with and without each property.
Section D. Seed augmentation.
Enumerating design pairs that the current knowledge base reasons about well.
An example seed of two quantitative variables produces a scatterplot pair one without binning and the other with binning. 
Another example seed of a quantitative and a discrete variables produces a dot plot pair with the discrete variables on the x axis and on the y axis.}]{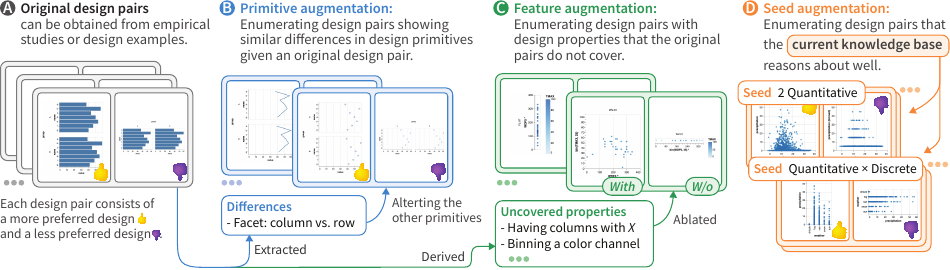}
  \caption{%
  Our work introduces data augmentation techniques for updating a visualization design knowledge base using example design pairs, in which one chart is deemed preferable to another.
  Given (A) an original set of design pairs obtained from empirical studies or example cases, (B) \textit{primitive augmentation} produces variations of the original designs by enumerating pairs that exhibit the same differences in low-level design primitives as an original design pair (\eg~the use of the \textit{x} channel); (C) \textit{feature augmentation} extends the original pairs by adding new pairs that exhibit high-level design features (\eg~binning on the \textit{color} channel) that the original pairs do not cover; lastly, (D) \textit{seed augmentation} enumerates design pairs that the current knowledge base reasons about well. %
  }
  \label{fig:teaser}
}

\graphicspath{{figs/}{figures/}{pictures/}{images/}{./}} % where to search for the images

\usepackage{tabu}                      % only used for the table example
\usepackage{booktabs}                  % only used for the table example
\usepackage{lipsum}                    % used to generate placeholder text
\usepackage{mwe}                       % used to generate placeholder figures

\usepackage{mathptmx}                  % use matching math font

\begin{document}

%%%%%%%%%%%%%%%%%%%%%%%%%%%%%%%%%%%%%%%%%%%%%%%%%%%%%%%%%%%%%%%%
%%%%%%%%%%%%%%%%%%%%%% START OF THE PAPER %%%%%%%%%%%%%%%%%%%%%%
%%%%%%%%%%%%%%%%%%%%%%%%%%%%%%%%%%%%%%%%%%%%%%%%%%%%%%%%%%%%%%%%

\firstsection{Introduction}

\maketitle

Visualization design knowledge bases encode conditions for making plausible and effective charts, often based on empirical findings. 
For instance, Draco~\cite{moritz2018:formalizing} has \textit{hard constraints}, violations of which result in `illegal' charts, and \textit{soft constraints} indicating preferences, inspired by visualization perception studies. 
An example of a hard constraint is using both log and linear scales for the same encoding channel. 
Soft constraints encode meaningful combinations of design choices, for example, `binning a quantitative variable on the \textit{x} channel' consists of several primitives (binning, data type, encoding channel). 
By assigning different weight terms to these preferences, knowledge bases can recommend charts with the lowest `cost' (sum of weights).

As our understanding of effective visualization design improves, a knowledge base needs to reflect up-to-date findings, such as updating weight terms for design features. 
However, revising a knowledge base can be challenging due to unexpected downstream effects, such as changes to chart rankings.
For example, manually updating weight terms may fail to account for their relationship to other weight terms.

To support knowledge base updates, prior work has proposed different methods.
For example, Schmidt~\ea~\cite{schmidt2024:dracova} propose a visual analytic tool for inspecting Draco.
Draco-Learn is a machine learning (ML) method to update weight terms using design examples used in empirical studies~\cite{moritz2018:formalizing}.
When updating weight terms either automatically or manually, however, relying on a few example designs may cause overfitting~\cite{zeng2024:tooManyCooks} and fail to assign appropriate weight terms to cases for which the original knowledge base was successful.

To improve feature coverage and avoid overfitting, different ML domains have adopted data augmentation techniques~\cite{mumuni2022:dataAug}, which modify original data points to generate an expanded and more reliable training dataset.
For instance, in computer vision, smaller training datasets (\eg~for object detection) have been augmented by adding rotated, cropped, or distorted images, achieving higher performance on cases outside of the training set. 
Such methods are not directly applicable to data visualization, as they may create invalid visualizations or modify their semantics. For example, cropping a section of a visualization might remove guides such as axes or legends.

We contribute augmentation techniques for data visualization, and apply them to update visualization knowledge bases.
We propose and compare three such techniques: \textit{primitive}, \textit{feature}, and \textit{seed} augmentation, as shown in \Cref{fig:teaser}.
Given input chart pairs from empirical studies, primitive augmentation generates new designs that preserve identified design differences among the input charts (\eg~use of a mark type, encoding combinations) to have balanced training data.
Next, knowledge bases reason based on higher-level design facts (\eg~use of a mark type with certain encoding channels), which we refer to as \textit{design features}.
Consequently, feature augmentation generates designs that assess design features that are insufficiently covered by the input corpus. 
Lastly, seed augmentation enumerates additional designs that the original knowledge base already reasons about well (\ie~those that exhibit low cost) to make an incremental update to the current knowledge base while reducing the risk of overfitting to new data points.

Our augmentation methods generate new chart examples, but can not reliably provide labels---namely, which chart to prefer in a given chart pair.
In contrast to image data where labels (\eg object names and locations) propagate through augmentations, visualization augmentation requires new labels.
To support the annotation process, we compare different methods for labeling augmented data: \textit{manual}, \textit{ML classifier}, \textit{active ML classifier}, and \textit{LLM} labeling.
Our automated approach trains a classifier on input design pairs to predict preferences for augmented pairs, while the active ML method also solicits manual labels for uncertain predictions using active learning methods.
Given advances in large language models (LLM)---and results showing that LLM chart recommendations (for value comparison tasks) reasonably align with prior studies~\cite{wang2025:dracogpt}---we also label augmented pairs using an LLM. 

We evaluate our methods in the context of learning updated Draco knowledge bases~\cite{yang2023:draco2,moritz2018:formalizing}.
In similar settings to the original Draco work---using a non-augmented chart corpus from Kim~\ea~\cite{kim2018:encodings}, Saket~\ea~\cite{saket2018:encodings}, and Mackinlay~\cite{mackinlay1986:APT}---our augmentation techniques lead to comparable or slightly improved performance ($\sim$97-98\%, measured using manual labels).
As expected, each augmentation technique improves performance when tested against corresponding augmented pairs.
When testing against the full data (both original charts from a larger corpus~\cite{zeng2023:dataset} and the augmented charts), feature augmentation (90\%) and the combination of all augmentations (89\%), outperform all other methods (75--81\%).
While manual labeling consistently produces the best results, automated methods are often competitive.
Classifier-based methods perform well for primitive augmentation examples, whereas LLM labeling provides more consistent accuracy levels for (often previously unseen) feature augmentation results.

To recap, our contributions are:
\begin{itemize}
    \item Three data augmentation techniques for generating visualization design pairs: primitive, feature, and seed augmentation.
    \item A range of labeling methods (from manual labeling, to classifier-assisted labeling, to LLM labeling) to scale annotation efforts.
    \item A quantitative evaluation of augmentation and labeling techniques when updating a Draco knowledge base.
    
\end{itemize}
All code, data, and results are available as Supplementary Material.

\section{Related Work}\label{sec:rw}
Our work is grounded in prior work in visualization design knowledge bases and data augmentation for machine learning (ML).

\subsection{Visualization Design Knowledge Base}\label{sec:rw:kb}
\noindent Prior work often represents visualization design principles using \textit{knowledge bases} and \textit{graphs}. 
First, a visualization design knowledge base encodes \textit{rules} for making \textit{grammatically correct} and \textit{preferred} charts.
Mackinlay's APT~\cite{mackinlay1986:APT} provides \textit{expressiveness} rules for visualizing all and only information available from data and \textit{effectiveness} rules for displaying it in a way that best supports rapid perception. 
For example, mapping a nominal variable to bar lengths is not \textit{expressive} because it implies non-existing ordinal relationships between categorical values.
Using a position encoding channel (\eg~\textit{x}, \textit{y}) is more \textit{effective} than using color or shape because human visual perception tend to be most sensitive to positional differences. 
Similarly, systems like BOZ~\cite{casner1991:boz}, Visage~\cite{schroeder1992:visage}, Vista~\cite{senay1994:vista}, SAGE~\cite{roth1988:sage1,mittal1998:sage}, and CompassQL~\cite{wongsuphasawat2016:compassQL} have rules for generating expressive and effective charts.
In addition to explicitly encoding relevant design criteria, visualization design knowledge bases function as reasoning modules for authoring tools with automated recommendation functionalities (\eg~Voyager~\cite{wongsuphasawat2016:voyager,wongsuphasawat2017:voyager}, Dupo~\cite{kim2023:dupo}).

Next, a visualization knowledge graph encodes similar information using \textit{relations} between different entities. 
For example, KG4VIS~\cite{li2022:kg4vis} and AdaVis~\cite{zhang2024:adavis} represent data and encoding characteristics as entities and encode their plausible and effective matches as relations (\eg~a quantitative variable to the \textit{x} channel).
To efficiently operate relatively large knowledge graphs, they use an embedding that represents entities and relations as numeric vectors.
Knowledge graph-based approaches aim to learn the scoring function defined on an embedding that estimates how effective each entity relation is.
To reduce the bias in training, these approaches enumerate unlikely entity relations (\ie~\textit{negative sampling}~\cite{sun2018:rotate}) by replacing entities with something else at each training iteration.
While our work focuses on knowledge bases, we later discuss implications of our approach to visualization knowledge graphs.

It is often difficult to update knowledge bases to reflect novel empirical findings because they embed rules within their reasoning systems.
For example, SAGE~\cite{mittal1998:sage} has a rule for generating a description for encoding channels that combines preconditions and operations. 
CompassQL~\cite{wongsuphasawat2016:compassQL} encodes its rules as modules that reason about design properties and their preferences together.
Instead, Draco~\cite{moritz2018:formalizing,yang2023:draco2} separates the design preferences from the design rules using \textit{weight terms}.
Specifically, Draco encodes design `features' as soft constraints associated with weight terms such that a certain feature can be preferred to others under given conditions. 
For instance, Draco can weight a feature for `binning' depending on the number of bins. 
In this way, one can update the knowledge base by updating the weight terms, which prevents costly restructuring of a knowledge base~\cite{falkner2013:kbchallenges}.

Rather than manually tune weights in an error-prone fashion, one can update weight terms in an automated fashion by training a classifier that predicts which chart in a design pair should be preferred to the other\cite{moritz2018:formalizing}.
In each pair, one chart is labeled as preferred to the other, for example, because a study finds it more effective (\eg more quickly or accurately decoded).
A regression-based classifier is trained on chart feature vectors, where each dimension is a soft constraint (design feature) and values are the count of violations of each constraint. 
The resulting model coefficients then provide a set of weight terms. 

Prior work has inspected visualization design knowledge bases, primarily Draco, through various methods. 
Schmidt~\ea~\cite{schmidt2024:dracova} developed a visual analytic tool for inspecting the relationship of Draco's constraints to help visualization practitioners understand how Draco works.
For example, the tool shows constraints shared and not shared by two candidate designs recommended by Draco.
However, manually updating Draco weight terms can be tedious and cause unexpected side effects relative to other constraints. 
To analyze chart recommendations from LLMs, Wang~\ea~\cite{wang2025:dracogpt} propose DracoGPT, an extension of Draco-Learn~\cite{moritz2018:formalizing} where an LLM agent compares pairs of visualization designs to provide a labeled training corpus.
Using a collection of 30 graphical perception studies~\cite{zeng2023:dataset}, Zeng~\ea~\cite{zeng2024:tooManyCooks} report that if a constraint appears too few times in the design pairs, then it can cause a knowledge base with updated weight terms to fail to generalize to other cases.
Under-generalization can stem from two sources: features that a knowledge base is not fully capturing, and weight terms overfit to too few cases.
By focusing on the latter, our work aims at providing a set of methods that augment example designs used for updating the weight terms of a knowledge base, using Draco as a case study.

\subsection{Data Augmentation}\label{sec:rw:da}
In machine learning (ML), a large amount of balanced training data is often critical to model performance. 
For ML-based computer vision tasks (\eg~object detection), researchers have applied data augmentation techniques to increase the volume and diversity of training data~\cite{mumuni2022:dataAug}.
Common augmentation techniques include geometric transformations (\eg~zooming, slicing, or rotating), photometric distortions (\eg~blurring, adjusting contrast, adding noise), region-level changes (\eg~swapping, replacing, and/or deleting regions in an image), and applying more complex transformations using deep-learning models.
Data augmentation techniques are useful when obtaining a large set of training data is infeasible or costly (\eg~medical imaging~\cite{chlap2021:medAug}, field agriculture~\cite{zou2024:agrAug}, chart object~\cite{li2022:structure}).
These techniques typically preserve labels for the data (such as object area annotations), as they retain or similarly transform the labels. 
For instance, adding noise to an image of a flower does not change the location and property of the target object. 

However, these augmentation techniques do not apply directly to learning visualization design knowledge.
For example, distorting or cropping a chart image is unrealistic for a visualization design process because this process will produce inexpressive charts. 
Furthermore, a slight change to a visualization design property, such as changing a mark type, may or may not affect perceptual effectiveness.
Data augmentation techniques for visualization design knowledge bases also need to make sure that the resulting data sufficiently cover the design characteristics that are represented by the knowledge base.
Models trained on insufficient data will fail to learn uncovered weight terms and may exhibit overfitting to less covered weight terms~\cite{zeng2024:tooManyCooks}.

Unlike traditional ML work, augmentation techniques for visualization can not propagate existing supervised labels, presenting the challenge of further labeling generated data.
For example, the same encoding choices can have different levels of effectiveness depending on mark types and data characteristics (\eg~scatterplot vs. line chart for time serial data~\cite{wang2018:line}).
Furthermore, we may not have enough information to automatically label augmented designs for insufficiently covered characteristics.
This challenge stems from the fact that the goal here is to learn preferences over design; this is analogous to learning the desirability of a photo, which existing image augmentation techniques do not fully address.
To effectively augment training data for updating a visualization knowledge base, our work proposes a set of augmentation techniques to provide improved coverage and compares different labeling strategies to annotate the augmented data.

\section{Background}\label{sec:bg}

\subsection{Draco and Answer Set Programming}\label{sec:bg:draco}
We develop our methods in the context of Draco (in particular, Draco 2~\cite{yang2023:draco2}), given its adoption in prior work and the formal reasoning capabilities offered by its use of Answer Set Programming (ASP)~\cite{lifschitz2008:asp,brewka2011:answer}.
ASP is a constraint-based programming method that expresses inference rules and constraints.
Given some initial facts, an ASP solver (\eg~Clingo~\cite{gebser2014:clingo}) enumerates candidates that are coherent with constraints.
In data visualization contexts, initial facts can include authors' intents about variables to visualize or preferred encoding channels.
Draco populates design properties applicable to authors' initial facts.

Draco expresses a visualization design using \textit{primitives} and \textit{features}. 
A primitive defines the existence and attributes of an entity, or element. 
For example, an \textit{x} encoding channel that maps a `profit' variable has the following primitives. 
First, \lstinline{entity(encoding,m,e0)}: an encoding channel \lstinline{e0} exists for a layer (or mark) \lstinline{m}.
Next, \lstinline{attribute((encoding,channel),e0,x)}: the encoding channel, \lstinline{e0}, has a \textit{channel} attribute set to \textit{x}.
Lastly, \lstinline{attribute((encoding,field),e0,profit)}:  the \lstinline{e0} encoding has a \textit{field} attribute set to \textit{profit}.

These primitives are too low-level to conceptually reason about a visualization design, so Draco defines meaningful features by combining relevant primitives.
For example, Draco has a design feature that captures if a layer, \lstinline{M}, uses a bar mark and has a continuous position channel and a discrete position channel:
\lstinline{preference(c_d_no_overlap_bar,M)}.
This feature includes primitives for mark type, encoding channels, and data types mapped to those channels.

Draco assigns numerical weight terms to those features to indicate relative preferences.
A negative weight encourages the corresponding design feature, whereas a positive value discourages it.
These preferences are technically called `soft constraints' in the sense that a visualization design can `violate' them, incurring the `cost' of the corresponding weight.
The violation of a soft constraint is not necessarily bad; instead it means that the design includes that feature.
Draco assigns the final cost to a candidate design by summing the weight terms for its design features. 
Assigning proper weights to soft constraints is an important maintenance job to keep Draco up-to-date with our understanding of---or subjective preferences for---visualization design. 

\subsection{Updating Draco}\label{sec:bg:updating}
\begin{figure}
    \centering
    \includegraphics[width=\linewidth, alt={A diagram for updating Draco from empirical studies.
There are six steps.
Step 1. Prepair design pairs.
An ordered pair has feature sets.
Step 1-A. Augment design pairs.
Step 1-B. Label with manual, ML, active ML, and LLM methods.
Step 2. PRocess training data. Feature sets are differed and converted to a vector.
Step 3. Train a model. A linear model of label vector Y equals coefficient matrix A times data matrix X.
Step 4. Evaluate.
The coefficient matrix A is converted to new weight terms, and using them predict the data again to get performance.}]{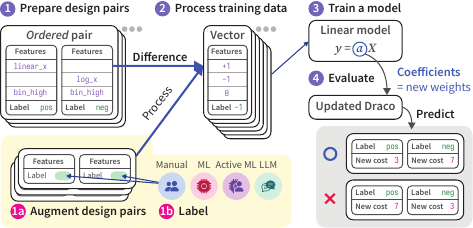}
    \caption{To update Draco, we (1) prepare designs into ordered pairs, (2) process them as feature difference vectors, (3)  train a linear model and obtain coefficients, and (4) use them as new Draco weight terms to predict the costs of design pairs. Augmentation and labeling (1a \& 1b) takes place before processing  (Step 2).}
    \label{fig:update:pipe}
\end{figure}

To update Draco weight terms using design examples, one needs to prepare ordered design pairs, featurize them, fit a model, and extract new weight terms, as shown in \Cref{fig:update:pipe}.
We assume a case where a researcher is using design pairs drawn from empirical studies and adding new pairs to an existing corpus. 

\bpstart{Step 1. Prepare design pairs} 
After empirical studies or discussions, we can obtain preferences about given designs.
For instance, an experiment may suggest a design is more effective than another. 
Then, those two designs form an \textit{ordered} pair where one is preferred to the other.

\bpstart{Step 2. Process training data} 
For each design in a pair, we collect its violated soft constraints (\ie~design features).
If a feature occurs $x$ times only in the preferred design in a pair, we mark it as $+x$.
If it happens $x$ times in the preferred design and $y$ times in the non-preferred design, we mark it as $x-y$.
If it has the same frequency (including zero occurrences) in both charts, we mark it as $0$.
In this way, we prepare the relative occurrences of design features for the input design pairs.

\bpstart{Step 3. Train a model}
We train an ML model that predicts which design in a pair is preferred given their relative occurrences of design features. 
Upon training linear models (\eg~support vector machine or logistic regression), we obtain coefficients for each feature.
We can apply those coefficients as new weight terms. 

\bpstart{Step 4. Evaluate a model}
We can evaluate the model by rerunning Draco with the updated weight terms. 
If the updated Draco model provides a smaller cost for the preferred design in a pair, we consider the result to be accurate. 
By applying this check over a test data corpus, we can compute an overall accuracy score. 

\bstartnc{When does augmentation happen?}
Data augmentation can take place after preparing design pairs (Step 1).
Our work aims to detect and fill gaps in the initial design pairs to train ML models in a more balanced way that enhances coverage and therefore generalizability.

\section{Augmentation Techniques}\label{sec:aug}

\noindent
Empirical studies usually focus on the impact of specific design choices, forming a relatively small set of design pairs to use in the above updating process.
For example, studies may include too few design stimuli, appearing only once or twice in the dataset, resulting in high sensitivity to those instances.
They may not fully cover the entire feature space of a knowledge base.
In addition, they might not include relatively simpler designs like a bivariate scatterplot or univariate histogram.
Relying only on such examples can be problematic, causing the updated knowledge base to underperform by excluding cases that are omitted or poorly covered by the design pairs.

To address this problem, we propose three data augmentation techniques for updating a visualization design knowledge base: \textit{primitive augmentation}, \textit{feature augmentation}, and \textit{seed augmentation}.
First, \textit{primitive augmentation} enumerates new pairs that preserve identified design differences found in existing training pairs (while then permuting other design aspects) to prevent the updated knowledge base from being overly sensitive to specific design feature combinations.
Next, \textit{feature augmentation} populates design pairs exhibiting design features (or combinations thereof) that are missing or occur infrequently in the existing training pairs. Here the goal is to explicitly increase feature coverage so that the resulting knowledge base can better generalize to a larger pool of designs.
Lastly, \textit{seed augmentation} generates chart pairs for which the current knowledge base is already confident of the result (\ie~low cost). 
The goal is to validate existing preferences and make incremental adjustments as needed, while also providing balance to reduce the risk of overfitting to additional augmentations.
We implemented each of these techniques relative to Draco.

A major challenge for our augmentation techniques is to effectively scope the enumeration to avoid generating visualization designs ``far'' beyond the original designs. 
For example, given a line chart pair, we could try a scatter plot or bar chart as reasonable extensions, yet a stacked bar chart layered with a scatter plot overlay exceeds the scope of the initial training pairs.
Pruning Draco's search space \emph{a priori} may cause unexpected downstream effects, making scoping tricky. 
For instance, if we set a limit for the number of \lstinline{entity}s in a design, some essential \lstinline{scale}s may not be generated. 
Our augmentation techniques thus incorporate scoping strategies in addition to generation methods. 

\begin{figure}
    \centering
    \includegraphics[width=\linewidth, alt={This figure has two sections: A and B.
Section A. Primitive-augmented pair.
Extract and preserve design differences.
The original pair consists of a line chart with a categorical color and log Y scale and a line chart with no color channel, a linear Y scale, and faceting over the categorical variable.
The differences are: color channel, row faceting, and log vs. linear scales.
The augmented pair has the same differences but the mark type is changed from line to point.
Section B. Feature augmented pair (binary). Minimize feature differences. 
Feature A: Bin-high
Feature B: C-D-Overlap-Area.
Feature A and B are ablated: without and with.
The without chart has binning on the Y axis. 
The with chart has a area mark.
Shared features are Interesting-Y, Value-Discrete-X, Linear-Scale, and so on.
Provoked features include Value-Point for the without chart and Value-Area for the with chart. }]{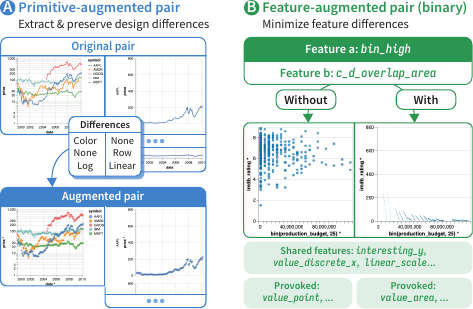}
    \caption{Primitive (A) and feature (B) augmentation examples.}
    \label{fig:aug:ex}
\end{figure}

\subsection{Primitive Augmentation}\label{sec:aug:design}
The primitive augmentation technique attempts to mitigate an updated model's sensitivity to a specific design case by repeating the design features covered by the training pairs with slight modifications.
First, given an original design pair, this technique first extracts the differences in design primitives, such as mark types, encoding channels, scale types, \etc 
Next, we remove design primitives that are shared by the designs in the pair, producing partial specifications. 
For each partial specification, we then run Draco to enumerate modified designs.
Lastly, we match the enumerated designs from each counterpart in a way that matched pairs exhibit the same design differences as the original pair. 
More specifically, this technique consists of the following steps.

\bpstart{(1) Extract design differences}
We extracted differences in design primitives between the charts in a given pair. 
This step links distributed design primitives by removing Draco's \lstinline{entity} identifiers, which is for the sake of ASP's logical reasoning.
For example, as shown in \Cref{fig:aug:ex}A, the original pair has a line chart with a log-scaled \textit{y} channel and a nominal color channel and a faceted line chart with a linear scale \textit{y} channel.
Draco expresses this information using multiple distributed primitives for each chart, connected by several entity identifiers.
Then, this step converts this into three abstract differences: (color, none), (none, row-facet), and (log, linear). 

\bpstart{(2) Form partial specifications}
From each chart in a pair, this technique removes design primitives except data-related ones (dataset, variables, their properties) and those appearing in the design differences captured by the previous step. 
Extending the earlier example, suppose that the charts share the same mark type, line (\lstinline{attribute((mark,type),m0,line).}), then we can remove this primitive from both charts. 

\bpstart{(3) Enumerate and match modified designs}
This technique uses Draco to enumerate complete chart designs for each partial specification.
Given that the partial specifications are already fairly complete (\eg~use of the \textit{y} encoding channel with a \textit{log} or \textit{linear} scale), Draco typically results in 10-15 designs for each (at most 30-40).
We also retain the number of layers and the number of encoding channels from original designs, which prevents a combinatorial explosion of the search space with unlikely designs. 
From the modified designs of each counterpart in a pair, we can form new pairs.
We run the design differences script (step 1 above) to evaluate these modified design pairs to check if they exhibit the original design differences and only those differences.

\subsection{Feature Augmentation}\label{sec:aug:feature}
The goal of our feature augmentation technique is to cover design features missing in the original training pairs to enhance the generalizability of the updated knowledge base. 
Given a design feature represented by Draco, this technique enumerates design pairs in which a counterpart includes that design feature while the other does not. 
First, we extract a partial specification and design features from the training design pairs.
Next, we obtain the list of the design features Draco represents but the training data do not exhibit (uncovered) or exhibit too few times (less covered).
Given a data specification and an insufficiently covered design feature, we use Draco to enumerate a set of designs that exhibit the feature (\textit{with} designs). 
For each \textit{with} design, we remove design properties, and then enumerate a set of designs that do not exhibit the feature (\textit{without} designs) in a way that they only differ from the \textit{with} design in terms of that specific feature.
This process results in new chart pairs that isolate design trade-offs not represented in the input training data.
We now detail each of these steps.

\bpstart{(1) Form partial specifications}
The first step is to gather basic information from each chart in a pair to form partial specifications for subsequent design enumeration. 
This step extracts information that can influence design features that Draco cannot simply enumerate, such as data-related facts: the dataset, the variables to use, and the characteristics of each variable (\eg~cardinality, entropy, \etc).
To prevent the enumeration process from exploding with too many unrealistic designs, we also include the coordinate system (Cartesian or polar), the number of layers, and the number of encoding channels.

\bpstart{(2) Obtain uncovered design features}
We run Draco for each chart in an original pair, while preventing it from adding any additional design primitives. 
This processing results in only the design features appearing in the chart. 
Then, we compute the frequency of design features and obtain those appearing less than a threshold value.
We set this threshold to seven, as Zeng~\ea~\cite{zeng2024:tooManyCooks} indicate that a trained output becomes less sensitive to rare features once they are repeated more than seven times. 

\bpstart{(3) Analyze design feature dependencies}
To effectively constrain the next enumeration step, this technique needs to be cognizant of the relationships between design features. 
For example, the \lstinline{aggregate} feature always occurs when \lstinline{aggregate_mean} occurs, but not conversely. That is, \lstinline{aggregate_mean} \textit{provoke}s \lstinline{aggregate}.
We do not want to test such commonly `provoked' features, as they already are tested via the `provoking' features. Formally, given two features $a$ and $b$, $(a \rightarrow b) \wedge \neg (b \rightarrow a) \Rightarrow (a \text{ provokes } b$).
On the other hand, the \lstinline{log_x} (using a log scale on \textit{x}) and \lstinline{linear_x} (using a linear scale on \textit{x}) features can not occur together because the scale of an encoding channel can not have multiple types; they are \textit{contradict}ing.
We use this information to avoid testing `contradicting' ($\perp$) features together because they will not enumerate any designs.
Formally, $(c \rightarrow a) \wedge (\neg c \rightarrow b) \Rightarrow a \perp b$, given two features $a$ and $b$, and a fact (primitive or feature) $c$.
The dependency analysis results in a directional graph where an edge $(a, \text{provoke}, b)$ indicates the provoking relation and $(a, \text{contradict}, b)$ marks the contradicting one.
This step prunes design features that are tested by or contradict other features.

\bpstart{(4) Enumerate ablation pairs}
Our feature augmentation technique essentially ablates a design feature given a partial specification obtained in the first step.
We use two ablation approaches: unary and binary. 
To consider the direct impact of a single design feature, the unary ablation looks at a single design feature as described above.
We can force a feature to appear by adding \lstinline{:- not preference(feature_name,_).} to a partial specification when running Draco. 
This constraint literally translates to `it must not be the case that the feature does \lstinline{not} appear.'
Similarly, we can prohibit a feature using \lstinline{:- preference(feature_name,_).}, meaning `it must not be the case that the feature appears.'
For each design feature, we select seven partial specifications and apply the ablation constraints.

We must also contend with cases where a feature behaves differently given the existence (or absence) of another feature.
For example, the \lstinline{bin_high} feature (the use of 12+ bins, resulting in a histogram) may not work well with the \lstinline{c_d_no_overlap_area} feature (using area mark for quantitative $\times$ discrete on \textit{x} and \textit{y}).
The binary ablation looks at a design feature given the presence and absence of another feature.
When we have two features, \lstinline{a} and \lstinline{b}, this technique enumerates two pairs: ablating \lstinline{b} given the presence of \lstinline{a} and ablating \lstinline{b} given the absence of \lstinline{a}.
Given that this tests \lstinline{b} across other features more than a few times, we set the threshold to be one for each pair of design features, resulting in one pair for either presence or absence of \lstinline{a}.
We also randomly select a partial specification for each pair of design features.  

\subsection{Seed Augmentation}\label{sec:aug:seed}
The seed augmentation technique aims to allow an updated knowledge base to support designs for which the current knowledge base already has high confidence, but empirical studies might not cover.
For example, within value comparison tasks, binning may be less preferred for a scatterplot, but preferred for a heatmap.
This may seem obvious, without requiring a perception experiment for confirmation.
However, if training pairs only include the heatmap case, the updated knowledge base may not reason well about the scatter plot case by overly preferring binning.
As opposed to building a knowledge base from scratch, our goal is to make an incremental update by testing cases believed to already be ``good'' and adjusting accordingly.
Assuming that the current design base works well with commonly used charts, the seed augmentation technique generates relatively simpler chart designs from a few curated partial specifications.
Here, a partial specification defines data variables to encode, the number of layers, and the number of encoding channels.
Given a partial specification for two quantitative fields, for example, Draco can generate a scatter plot with or without binning one of them, as shown in \Cref{fig:teaser}D.

The first step is to run the Draco solver to enumerate many single chart designs given a data specification.
To generate simpler, lower-cost charts, we limit the total cost per chart. 
From the enumerated designs, we select the top $N$ designs with different costs (\ie~lowest total costs) from the enumerated designs.
We then couple the selected designs into pairs, resulting in $N(N-1)/2$ pairs per data specification.

\section{Labeling Techniques}\label{sec:labeling}

\noindent 
As mentioned earlier, these augmentation techniques modify original visualization designs, and the labels for a pair (which chart is preferred) may not persist across augmentations.
Thus, we need to label them for training and test purposes.
One approach is to label them manually, for example via expert judgments.
However, these augmentation techniques can result in a few thousand additional pairs, and so are tedious to manually label.
Thus, we also propose (semi-)automated labeling methods in addition to manual labels, as even noisy labels may benefit learning~\cite{ratner2017:snorkel,wang2025:dracogpt}.
We apply these labeling techniques to the primitive and feature augmentation methods.
The seed augmentation uses the current knowledge base's output, as its purpose is to preserve a degree of balance with the current state.

\bpstart{Classifier-based labeling}
First, we can train a classifier based on the original pairs and predict the preferences of augmented pairs.
The training data for this method consists of design primitive difference vectors of original pairs; this differs from the standard Draco-Learn approach, which focuses on design features (soft constraints), not lower-level primitives.
For example, Draco encodes a log-scaled, quantitative \textit{color} channel as a set of design primitives: \lstinline{entity(encoding,m,e0)}, \lstinline{attribute((encoding,channel),e0,color)}, 
\lstinline{attribute((encoding,field),e0,f)}, 
\lstinline{entity(field,root,f)}, 
\lstinline{attribute((field,type),f,number)}, 
\lstinline{entity(scale,root,s0)}, 
\lstinline{attribute((scale,channel),s0,color)}, and 
\lstinline{attribute((scale,type),s0,log)}.
Here, the last term in \lstinline{entity} and the second term in \lstinline{attribute} are element identifiers for the internal use of Draco. 
To not duplicate the same primitive, we remove those identifiers and map directly to visualization elements.
The above example is converted to: \lstinline{color}, \lstinline{color.quantitative}, and \lstinline{color.log}.
We then generate a one-hot feature vector for each chart and take the difference of the feature vectors for a given pair, forming a training data point. 
Next, we can also apply active learning methods~\cite{settles2009:aml} in which a classifier is initially trained on both original and augmented pairs without labels. 
Then, we iteratively query data points uncertain to the classifier and provide labels to those only.

\bpstart{LLM-based labeling}
Lastly, we can label pairs using a large language model (LLM) given their reasonable performance in predicting pairwise chart preferences for value comparison tasks~\cite{wang2025:dracogpt}.
We use Wang~\ea~\cite{wang2025:dracogpt}'s DracoGPT method for LLM-based labeling, listing design primitives in a JSON format without Draco object identifiers.

\section{Evaluation}\label{sec:exp}

We evaluate our augmentation techniques and compare labeling methods in an experiment producing updated Draco weights.
The evaluation assumes a setting where an individual visualization researcher is trying to update Draco for their own purposes; rather than producing a ``comprehensive'' Draco instance whose full training data is rigorously grounded in empirical performance data. 
Hence we evaluate against expert judgments, rather than experimental results, when assessing performance.
We focus on learning weights for Draco design features relating to \lstinline{value} tasks (\ie~learning preferences for comparing individual values, rather than aggregate or summary properties).
Specifically, we aim to understand how the different augmentation techniques affect the performance of updated knowledge bases. 

\subsection{Methods}\label{sec:exp:methods}

\subsubsection{Initial Chart Pair Data}
We use design pairs from 30 graphical perception studies catalogued by Zeng~\ea~\cite{zeng2024:tooManyCooks}.
In particular, this dataset includes studies with broader design coverage (Mackinlay~\cite{mackinlay1986:APT}, Kim~\ea~\cite{kim2018:encodings}, and Saket~\ea~\cite{saket2018:encodings}) as well as those looking at more specific design cases (\eg~time-series charts~\cite{gogolou2019:time}, small multiples~\cite{ondov2019:multiples}).
Accordingly, we group the three larger-scale studies into a \textbf{Baseline} dataset, and the other 27 studies into an extended \textbf{Zeng+} set.
Together, these datasets comprise 786 pairs (696 \textbf{Baseline} and 90 \textbf{Zeng+}) for the \lstinline{value} task (\Cref{tab:dataset}).

\begin{table}[t]
\caption{Design pairs by data groups, augmentation techniques, and test splits. We used 5-fold cross validation and 15\% holdout validation set for the evaluation. \textbf{Baseline} pairs are from Kim~\ea~\cite{kim2018:encodings}, Saket~\ea~\cite{saket2018:encodings}, and Mackinlay~\cite{mackinlay1986:APT}, and \textbf{Zeng$^+$} pairs are from the collection by Zeng~\cite{zeng2023:dataset}.}\label{tab:dataset}
\centering
\includegraphics[width=\linewidth]{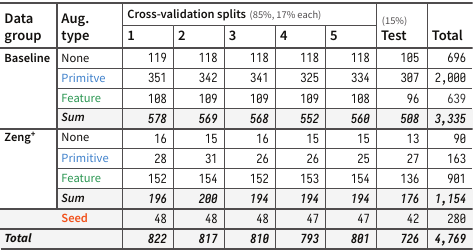}
\end{table}

\subsubsection{Augmentations}

\begin{figure}[t]
\centering
\includegraphics[width=\linewidth, alt={This figure has three bar charts: A, B, and C.
A. Primitive. N is 2163.
41 pairs had one enumerated pairs.
246 pairs had two enumerated pairs.
161 pairs had three enumerated pairs.
161 pairs had four enumerated pairs.
80 pairs had five enumerated pairs.
9 pairs had six enumerated pairs.
7 pairs had seven enumerated pairs.
B. Feature-Unary. N is 425.
2 pairs had six enumerated pairs.
59 pairs had seven enumerated pairs.
C. Feature-Binary. N is 1115.
143 pairs had one enumerated pairs.
486 pairs had two enumerated pairs.}]{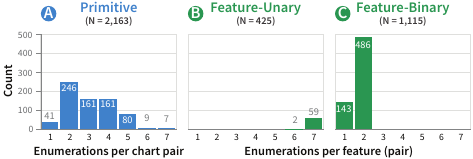}
\caption{Enumerated cases per original design pair for (A) primitive, (B) unary feature, and (C) binary feature augmentation.}\label{fig:augdist}
\end{figure}

\bpstart{Primitive augmentation}
For each original design pair, we enumerated up to seven additional pairs, resulting in 2,692 pairs.
While Draco's hard constraints pruned out a large set of \textit{grammatically incorrect} charts (\eg~a log-scaled categorical encoding), we found remaining \textit{visually inexpressive} charts. 
These charts are illegible due to excessively high graphical density (\eg~hundreds of overplotted line segments, overlapping bars without any aggregation).
We manually removed 529 such pairs while labeling them, resulting in 2,000
new pairs derived from the \textbf{Baseline} group and 163 new pairs from the \textbf{Zeng$^+$} group.

Designs from the \textbf{Baseline} group tend to exhibit minimum design differences (\eg~the use of a log scale).
In contrast, those from the \textbf{Zeng+} group tend to contain more design differences, which limits the space of design variations.
For example, Javed~\ea~\cite{javed2010:timeseries} compared mark types (line \& area) and layout options (faceting \& layering) in time series charts with different layouts, resulting in zero augmentations.
\Cref{fig:augdist}A displays the distribution of primitive-augmented designs.

\bpstart{Feature augmentation}
From the original input design pairs, we obtained 38 unique data specifications and 61 features that were missing or insufficiently covered (appearing less than 7 times).
For unary ablation, we initially enumerated 427 pairs (seven per feature) by randomly choosing different data specifications for each pair.
After removing two illegible pairs (as we did for primitive-augmented pairs), we used 425 unary ablation pairs.

After unary ablation, we analyzed patterns of design feature co-occurrence in the original and unary ablated charts, in addition to a static analysis of provoking and contradictory relationships between features (details in Supplementary Material).
We excluded feature pairs exhibiting the following characteristics: (1) commonly appearing in charts (more than 90\% of charts) and (2) feature pairs that are contradictory or provoking, as described earlier.
For example, the \lstinline{horizontal_scrolling_x} (mapping a high cardinality field to the \textit{x} channel) and \lstinline{high_cardinality_shape} (mapping it to the shape channel) features can not happen together.
The former requires a discrete or binned field to be mapped to the \textit{x} channel while the latter requires it to be the shape channel, but our data specs typically have a single discrete channel or binned quantitative channel. 
This step resulted in 315 ordered feature pairs (say $a$ and $b$), from which we enumerated 1,129 pairs (1-2 \textit{with}-$a$ pairs and 1-2 \textit{without}-$a$ pairs) also with random data specifications. 
We then removed 14 illegible pairs, resulting in 1,115 pairs.
By enumerating four chart pairs per feature pair, the ablation of each feature appears more than 7 times. 
\Cref{fig:augdist}B and C shows the distribution of the number of feature-augmented designs.

\begin{table}[t]
\caption{Data specifications for the seed augmentation technique by variable types.}\label{tab:data-spec}
\centering
\includegraphics[width=\linewidth]{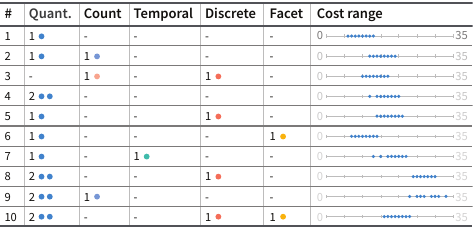}
\end{table}

\bpstart{Seed augmentation}
We generated the top eight charts from each of ten data specifications we curated to cover common use cases (\Cref{tab:data-spec}), forming $280$ pairs ($=10 \times \binom{8}{2}$).
They represent univariate, bivariate, time-series, and categorical cases with positional channels and faceting.
We assume that Draco is already effective in predicting the preferences of those common use cases, so use Draco's existing weights to label the pairs.
These specifications result in relatively simple chart designs including histograms, scatterplots, line graphs, and faceted charts.
We enumerated 600 charts per data specification by limiting the number of design features up to 20. 
The chart costs ranged from 6 (for a chart with a single quantitative variable) to 64 (for one with two quantitative variables). 
Cognizant of self-training bias, we used only a small portion of seed-augmented pairs (less than $6\%$ of the entire pairs).

\subsubsection{Labeling}
We labeled the augmented designs in the following ways.
For the seed-augmented cases, we used Draco 2~\cite{yang2023:draco2} costs (single labeling method). 
For the primitive- and feature-augmented cases, we used \textbf{manual}, \textbf{ML classifier}, \textbf{active ML classifier}, and \textbf{LLM} labels. 
First, a visualization expert (the first author) iteratively labeled pairs using a few heuristics: preferring fewer data transformations (aggregation, log, binning), point or tick marks, and positional encoding channels.
Next, we trained Support Vector Classifier (SVC) and Multi Layer Perceptron (MLP) models for automated labeling, and chose MLP models which achieved about 77.5\% of 5-fold cross-validation accuracy (average) with the original pairs (SVC achieved below 60\%).
Then, we applied active learning by querying 20 pairs per iteration (total of 20 iterations) using the same MLP model, achieving 79\% average CV accuracy. 
Lastly, we applied the Draco GPT method~\cite{wang2025:dracogpt} to collect LLM labels. 
We used OpenAI's GPT-4o-mini model given its low cost and its comparable performance to the models used in the original DracoGPT paper.
LLM labels exhibited 57.47\% agreement with the original \textbf{Baseline} design pairs.
In contrast, Wang~\ea~\cite{wang2025:dracogpt} found 79.49\% label agreement using GPT4 and the Kim~\ea~\cite{kim2018:encodings} data only.
LLM labels achieved 72.87\% agreement with the \textbf{Zeng$^+$} pairs.

\begin{table*}[t!]
\caption{Prediction accuracy for logistic regression models trained and tested on the original and augmented data (top: \textbf{Baseline} data, bottom: \textbf{Zeng$^+$} data). Linear SVC models showed similar performance at large, refer to Supplementary Material for details.}\label{tab:res:LR}
\centering
\includegraphics[width=\linewidth]{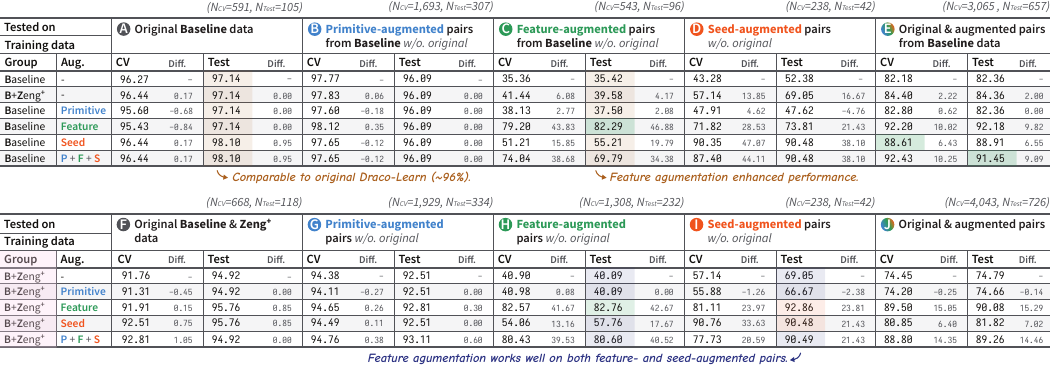}
\end{table*}

\subsubsection{Models}
\begin{figure}[t]
\centering
\includegraphics[width=\linewidth, alt={A combination for model configurations.
Two model types: Linear SVC and Logistic regression.
Two data groups: Baseline and Baseline + Zeng+
Five feature augmentation options: None, Manual, ML, Active ML, and LLM.
Five primitive augmentation options: None, Manual, ML, Active ML, and LLM.
Two seed augmentation options: None and Included.
This results in 200 models.
An example model is: SVC on Baseline data with LLM-labeled feature augmentation, active ML-labeled primitive augmenation, and seed augmentation included.}]{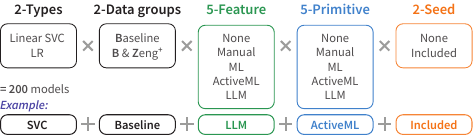}
\caption{Configuration for models for training and an example for how to read a model label. Linear SVC (support vector classifier with a linear kernel). LR (logistic regression).}\label{fig:models}
\end{figure}

To learn updated Draco weigh terms, we train either support vector classifier (SVC, with a linear kernel) or logistic regression (LR) models, and extract the fitted coefficients.
At a high-level, a model $f$ takes as input a pair of charts $\mathbf{a}$ and $\mathbf{b}$ and predicts a relative preference:
\begin{equation}
    f(\mathbf{a}, \mathbf{b}) =
    \begin{cases}
    -1 & \quad \text{if } \mathbf{a} \text{ is preferred}\\
    0 & \quad \text{if } \mathbf{a} \text{ and } \mathbf{b} \text{ are similar}\\
    1  & \quad \text{if } \mathbf{b} \text{ is preferred}
  \end{cases}
\end{equation}
For each model type, we permuted the data groups (Baseline \& Zeng$^+$), the augmentation methods (seed, primitive, \& feature), and the labeling methods (manual, ML, active ML, \& LLM), as shown in \Cref{fig:models}.

\subsubsection{Data preparation}
Given the above model, the training data consist of $(\mathbf{a}, \mathbf{b}, y)$ triples.
Given a design pair, $\mathbf{a}$ and $\mathbf{b}$ are the vectors of the counts of design features appearing in the charts and $y$ is the preference label ($-1$, $0$, or $1$) for the pair.
To balance the prediction categories, we duplicate each pair by rotating their orders: $(\mathbf{a}, \mathbf{b}, y)$ and $(\mathbf{b}, \mathbf{a}, -y)$.
The difference between chart vectors ($\mathbf{a} - \mathbf{b}$) is fed to a model.
After sampling 15\% for a holdout test set, we split the remaining training data into five splits for cross-validation (\Cref{tab:dataset}).
Because some models (\eg~logistic regression) only perform binary classification, we further transform pairs labeled as `equal' (by human) or contradictory (by LLM) by duplicating and swapping labels to cancel them out: $(\mathbf{a}, \mathbf{b}, 1)$, $(\mathbf{a}, \mathbf{b}, -1)$, $(\mathbf{b}, \mathbf{a}, 1)$, and $(\mathbf{b}, \mathbf{a}, -1)$.
We duplicate the pairs after determining cross-validation splits so that the duplications belong to the same split.

\subsubsection{Measuring Performance}
After training a model, we use the coefficients (representing design features) as updated weight terms.
As ASP only allows for integer weight terms, we normalize and round the coefficients to the range  $[-1,000, 1,000]$ preserving signs and precision.
For example, if the maximum coefficient is $1.223$ and the minimum is $-0.712$, the integer weights range from $-712$ to $1,223$.
Preserving the signs of learned coefficients is important, as negative weights represent explicitly `preferred' design features.
By applying the updated weight terms, we can compute the new costs for the charts in each test pair.
If the cost of a preferred chart is lower (or vice versa), we consider this test pair to be \textit{compliant}. 
When the charts in a pair are deemed similar (\ie~label of $0$), then it is marked as compliant when the difference in cost is less than or equal to 2.
In our testing, there were only three primitive-augmented and six feature-augmented pair labeled as 0.
Across our models, they all resulted in the cost difference of 2 or less. 
Because we duplicate the training pairs, models make two (or four) predictions. 
If the predicted compliance of a pair is not consistent, then we consider it \textit{non-compliant}. 
The prediction accuracy is computed as the number of compliant pairs over the number of total pairs.

\subsection{Results}\label{sec:exp:res}

\subsubsection{Baseline + Augmentation (Train) $\rightarrow$ Baseline (Test)}
We first compare augmentation techniques (with manual labels) and the addition of the \textbf{Zeng$^+$} data in terms their performance on both original and augmented pairs from the \textbf{Baseline} data.
We provide the full performance scores in Supplementary Material.

\bstartnc{Tested only on the original \textbf{Baseline} data} (\Cref{tab:res:LR}A), our models generally achieved 95\%-97\% average accuracy in cross validation and 95-98\% on the holdout test set.
This result is comparable to the original Draco-Learn~\cite{moritz2018:formalizing} (about 93\%-96\%).
We compare the performance of models using the \textbf{Zeng$^+$} dataset and augmentation techniques (using manual labels).
Linear SVC and Logistic Regression models trained only on the \textbf{Baseline} data achieved about 97-98\% test accuracy. 
The addition of the \textbf{Zeng$^+$} pairs and \textbf{augmented} pairs did not change performance drastically, which confirms that our augmentation techniques did not ``harm'' the update process.

\bstartnc{Tested on the augmented data},
learned Draco models exhibited similar performance for the \textbf{primitive-augmented} pairs whereas they showed more drastic differences for the \textbf{feature-} and \textbf{seed-augmented} pairs. 
The stable performance on primitive-augmented pairs (96\% test accuracy, \Cref{tab:res:LR}B) makes sense because this technique amplifies features that the Baseline pairs already have.
In contrast, models \textbf{without feature augmentation} performed poorly ($\sim$50\% or lower, \Cref{tab:res:LR}C) while feature augmentation achieved 82\% test accuracy with logistic regression.
Likewise, models trained using \textbf{seed-augmented} pairs performed well on the corresponding augmented test pairs (\Cref{tab:res:LR}D). 
Interestingly, models trained on the \textbf{feature-augmented} pairs performed decently well on the seed-augmented pairs (73\% test accuracy).
Across models, combining other augmentation techniques did not lower the performance significantly. 

\bstartnc{When testing models on the original and augmented \textbf{Baseline} pairs},
we again observed that models trained on \textbf{feature-augmented} pairs and those trained with all three techniques improved the performance by 9.09\%, relative to non-augmented models (\Cref{tab:res:LR}E).

\subsubsection{Baseline + Zeng$^+$ Augmentation (Train) $\rightarrow$ All (Test)}
Next, we compare the performance of the \textbf{augmentation techniques} (manually labeled) on the entire original and augmented pairs.

\bstartnc{Tested on the original \textbf{Baseline} and \textbf{Zeng$^+$} data} (\Cref{tab:res:LR}F), models generally achieved 92\% CV accuracy and 95\% test accuracy.
Models trained on \textbf{feature-augmented} pairs (in addition to the original pairs) performed slightly better when tested on the original pairs by 1.75\% using Linear SVC (94.9\% $\rightarrow$ 96.6\%). 

\bstartnc{Tested on the augmented \textbf{Baseline} and \textbf{Zeng$^+$} data}, models exhibited consistent patterns as observed earlier.
Models generally had similar performance on the \textbf{primitive-augmented} pairs (\Cref{tab:res:LR}G), while using all three techniques achieved the highest performance (up to 2.4\% increase). 
When tested on the \textbf{feature-augmented} pairs (\Cref{tab:res:LR}H), models using feature augmentation achieved reasonable performance.
When tested on the \textbf{seed-augmented} pairs (\Cref{tab:res:LR}I), models showed 55-90\% CV accuracy and 67--93\% test accuracy.
Models trained on the \textbf{seed-augmented} and \textbf{feature-augmented} pairs performed better than others (by 21--26\%).

\bstartnc{Tested on the entire data}, models showed 74--90\% CV and test accuracy (\Cref{tab:res:LR}J).
Models trained on \textbf{feature-augmented} pairs improved performance by 12-15\%, and models trained using \textbf{seed-augmented} pairs also showed higher performance by 6--7\%, relative to non-augmented models. 

In sum, our augmentation techniques generally improve the performance, with feature augmentation being most helpful overall.
Each augmentation technique showed higher performance when testing corresponding augmented pairs, while using all augmentation techniques largely preserved the individual techniques' performance.

\begin{table}[t!]
\caption{Prediction accuracy for logistic regression models using different labeling techniques for training data. Human labels were used for scoring.}\label{tab:res:label}
\centering
\includegraphics[width=3.3in]{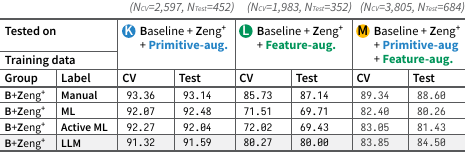}
\end{table}

\subsubsection{Labeling Techniques}
Next, we compare the labeling techniques (manual, ML, active ML, and LLM).
We use those labeling techniques for training and score predictions based on manual labels.
Relative to manual labels, ML labels agree 78.11\%, active ML labels agree 77.72\%, and LLM labels agree 73.66\%.
Although these labeling techniques under-perform manual labeling, they exhibit complementary strengths.
Models trained on the ML and active ML labels have comparable performance to manual labels for the \textbf{primitive-augmented} pairs (about 4\% lower than manual, \Cref{tab:res:label}K), but perform poorly when tested on \textbf{feature-augmented} pairs.
On the contrary, \textbf{LLM} labels under-perform for \textbf{primitive-augmented} pairs, but have higher prediction accuracy when the test data includes \textbf{feature-augmented} and \textbf{Zeng$^+$} pairs (\Cref{tab:res:label}L). 
This performance discrepancy may stem from the classifier (ML) labeler is more tightly optimized for the original data (which primitive augmentations hew to), while LLMs draw on a much larger training corpus (better supporting prediction on more varied, unseen designs). 
These methods have balanced performance on the entire data (\Cref{tab:res:label}M).

\subsubsection{Ad-hoc Analysis}
From the analysis, we observed interesting patterns. 
First, augmentation techniques in general performed well on their targeted test pairs without harming each other's behavior.
Next, models with and without \textbf{feature augmentation} showed the most significant performance gaps.
Third, \textbf{primitive augmentation} only marginally improved performance, potentially because non-augmented models already achieve high performance using the original training data.
Lastly, automated labeling techniques exhibited complementary performance.
To further inspect these patterns, we conducted ad-hoc analyses examining shifts in weight terms, cosine similarity between data groups, and training on heterogeneous pairs.

\begin{figure}[t]
\centering
\includegraphics[width=\linewidth, alt={There are five bar charts labeled A to E.
From A to D, gray bars indicate when augmentation pairs are added to original pairs; and yellow bars indicate when original pairs are added to augmentation pairs.
Bars can be directed to negative or positive. The value ranges from minus 50 to 50.
A. Weight shift due to primitive augmentaiton.
The bar heights are low in general but yellow and gray bars have opposite directions. 
B. Weight shift due to feature augmentation.
Bar heights are high in general with yellow and gray bars in opposite directions. Yellow bars are generally higher than gray bars.
C. Weight shift due to seed augmentation.
Yellow bars are in general higher than gray bars, and they tend to be in opposite directions.
D. Weight shift due to primitive, feature, and seed augmentations. 
Grat bars are in general low and evenly distributed, while yellow bars are sparse and higher.
E. Weight shift of ML and LLM labels from manual for primitive and feature augmentation techniques. 
Teal bars are for ML labels and yellow bars are for LLMs. 
They are in general in a similar directions but yellow bars had shifts that teal bars did not have.}]{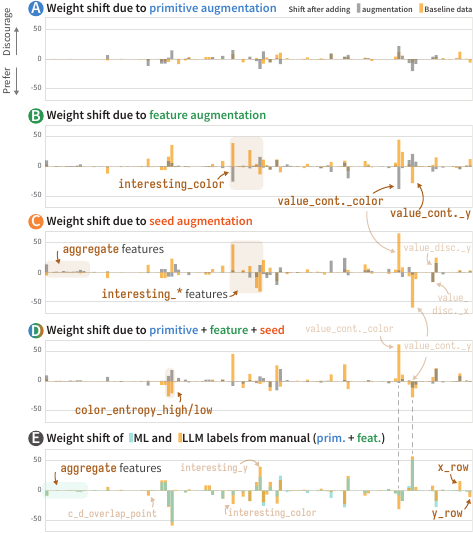}
\caption{Shift in weight terms by adding augmentation techniques and Baseline data (A-D, with manual labels), and by using ML and LLM labels compared to manual labels. 
The weight term differences are multiplied by the relative frequencies of corresponding features in the training data.
} \label{fig:weight-shift}
\end{figure}

\bpstart{Weight term shift}
To see what our models learned through augmentation techniques and labeling methods, we compare shifts in learned weights of models trained on different augmentation techniques and labeling methods (\Cref{fig:weight-shift}).
We additionally trained models using only augmented pairs (\ie~without the original Baseline pairs) to see how weight terms shift in both directions (by adding augmentation and by adding the original pairs). 
In \Cref{fig:weight-shift}A, each yellow bar represents the change to the corresponding feature's weight term from the model trained only on the augmented pairs (Primitive) to one including the original pairs (Primitive + Baseline). 
Each gray bar indicates the weight term change from the model trained only on the original pairs (Baseline) to one including the augmented pairs (Primitive + Baseline). 
The bar heights are weighted by their frequency in the entire training data. 
\textbf{Primitive augmentation} did not shift much compared to the Baseline data (\Cref{fig:weight-shift}A) while \textbf{feature augmentation} made a larger shift in quite opposite directions (\Cref{fig:weight-shift}B).
This helps explain the relatively lower performance on the feature-augmented pairs.
Next, our seed data specifications did not include \lstinline{interesting_*} (using a field of interest on a certain channel), so the model with the Baseline pairs learned about them (\Cref{fig:weight-shift}C). 
Plus, a large portion of the Baseline data is from Kim~\ea~\cite{kim2018:encodings}, which compares color, \textit{x}, and \textit{y} frequently, so using the Baseline pairs learned related features (\lstinline{value_continuous_*}). 
Using all three techniques exhibits regularization effects on the weight term shift (\Cref{fig:weight-shift}D). 

\newpage

Weight terms learned from LLMs and ML classifier labels shifted in a similar direction compared to those learned from manual labels, but to different degrees (\Cref{fig:weight-shift}E).
LLM labels showed bigger shifts as they had lower agreement with manual labels.
While relative frequencies are low, models on those labels learned some features in different directions. 
The Baseline model trained on ML-labeled pairs (teal) shifted more in the aggregate features, while the model trained on LLM-labeled pairs (yellow) learned more about layout-related features like \lstinline{c_d_overlap_point} (using point mark for continuous $\times$ discrete channels on \textit{x} and \textit{y}), \lstinline{x_row} (using the \textit{x} channel and row facets), and \lstinline{y_row} (using the \textit{y} with row facets).

\begin{figure}[t]
\centering
\includegraphics[width=\linewidth, alt={This is a heatmap matrix for cosine similarity between data sets and augmentation techniques for Seed, Baseline original, Baseline with Primitive, Baseline with Feature, Zeng+ original, Zeng+ with Primitive, and Zeng+ with Feature. 
They are mostly range from -0.05 to 0.05, with Baseline original and Baseline with Primitive with near 0.2 cosine similarity (highest).}]{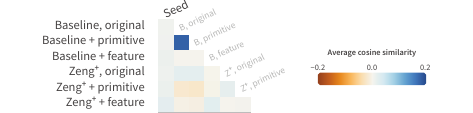}
\caption{Average design feature cosine similarity between data and augmentation types. A similarity of zero means two data sets are orthogonal.} \label{fig:cosine}
\end{figure}

\bstartnc{Cosine similarities between data groups and augmentation types} were generally low, ranging from $-0.02$ to $0.17$ (\Cref{fig:cosine}), which aligns with the performance discrepancies across different test sets. 
While generally low cosine similarity can be attributed to a relatively large set of features in our training data (126), most data groups had average cosine similarities from $-0.02$ to $0.02$, indicating their distinctiveness.
Given that data groups and augmentation techniques tend to focus on different design features, models with different configurations may have learned weight terms in complementary ways, implying the need for using these techniques together.
The low similarities correspond to increased coverage of the represented design space.

\newpage

\bstartnc{What if there is no Baseline data?}
We additionally trained models using only the original and augmented \textbf{Zeng$^+$} data. The goal was to assess the impacts of augmentation techniques when the original pairs are orthogonal.
First, models trained on the original \textbf{Zeng$^+$} data result in only 54-61.5\% test accuracy. 
\textbf{Primitive augmentation} improved the performance on both original and primitive-augmented data by 7\% (Linear SVC, 54$\rightarrow$61\% for the original, and 63$\rightarrow$70\% for the primitive-augmented pairs).
Models trained on the original \textbf{Zeng$^+$} data showed below 40\% performance on the \textbf{feature-augmented} pairs, while models using the feature augmentation resulted in 85\%.
\textbf{Seed augmentation} improved the performance on the original data, resulting in 77\% test accuracy (Linear SVC and logistic regression).
Using all three techniques resulted in the highest test accuracy on the original and augmented \textbf{Zeng$^+$} pairs (80\%, logistic regression). 
This result further demonstrates the value of data augmentation, including in relatively ``data poor'' settings such as the \textbf{Zeng$^+$} data alone.

\section{Discussion}\label{sec:discussion}
Using augmentation techniques, our models generally achieved reasonable performance.
For the \textbf{Baseline} pairs (from Kim~\ea~\cite{kim2018:encodings}, Saket~\ea~\cite{saket2018:encodings}, and Mackinlay~\cite{mackinlay1986:APT}), our models showed comparable performance given that prior work has already high performance with overall 95\% of pairwise prediction accuracy.
By adding augmented chart pairs, our models showed reasonable performances around 70-90\%. 
To compare, prior approaches to visualization recommendations (\eg~VizML~\cite{hu19:vizml} and LQ$^2$~\cite{wu21:lq2}) provide around 75-80\% prediction accuracy using more advanced non-linear models on fewer design features.
Machine learning for visualization is difficult because the effectiveness of a visualization design may not reside in a monotone, linear space.
Given that, our primitive augmentation technique, which is closer to the sense of ``tweaking'' in object detection, supported predicting diverse cases with reasonable performance. 
Next, by generating ``unseen'' and ``already good'' examples, our feature and seed augmentation techniques allowed us to update a design knowledge base to novel cases without harming the cases it is already good at. 
To make this approach scalable, automated labeling techniques showed reasonable performance relative to manual labels.
We now discuss implications to future research for supplementing and extending our approaches.

\bpstart{Applying augmentation}
At a high level, we recommend using primitive augmentation for keeping the balance, and seed and feature augmentation techniques for generalization.
When updating knowledge bases using fewer or highly orthogonal original pairs, enumerating primitive-augmented pairs will be important as shown in the above ad-hoc analysis, aligned with findings from Zeng~\ea~\cite{zeng2024:tooManyCooks}.
For labeling techniques, we suggest using an ML classifier for primitive augmentation and LLM labels for feature augmentation, as an ML classifier trained on the original data is less likely to generalize to unseen cases. 

\bpstart{Limitations}\label{sec:discussion:limitations}
Our main goal was to understand how three augmentation techniques influence updated weights in Draco.
We do not claim that manually-labeled design pairs in our evaluation are some kind of ``golden'' set. 
These labels can be understood as representing preferences that a developer or researcher wants to impart to a knowledge base.
Next, we only looked at Draco's soft constraints for the `value' task; future work could extend our approach to other task types. 
Lastly, we did not explicitly consider the discrepancy in effect sizes from different empirical studies, a limitation shared with prior work~\cite{moritz2018:formalizing, wang2025:dracogpt, yang2023:draco2, zeng2024:tooManyCooks}.

\bpstart{Automating parts of visualization research}
While we do not claim that we can replace human-subject perception studies with automated techniques, we highlight interesting aspects to consider in (partially) automating visualization science in terms of assessing research problems, aligning research outcomes, and supplementing software development.
First, empirical visualization studies tend to involve an exploration stage where researchers compare different study stimuli. 
At this stage, our feature augmentation techniques can support looking at examples that prior research has not looked at yet, or what a pilot study outcome is missing in a larger context. 

Second, when building large-scale visualization software, developers will want to include ``good'' default choices based on evidence.
However, there are gaps in the empirical record, some of which may have little immediate research value (as opposed to, say, engineering value). Plus, it is challenging to integrate a large body of existing evidence. 
Our augmentation techniques with automated or mixed labeling methods can provide developers with a reasonable starting point in a scalable way, enhanced by human interventions to a degree.

\bpstart{Use of LLMs for Augmentation}
We used LLMs to label the augmented designs, but not for generation, as we aimed to develop principled and explicit techniques.
For example, our feature augmentation method generates design pairs that precisely ablate a certain feature in a given knowledge base. 
LLMs cannot guarantee the same level of sensitivity due to stochastic behavior.
Similarly, LLMs may tweak design properties that need to be preserved for the primitive augmentation.
In addition, LLMs may not generate complete design specs given partial specifications, as reported by Wang~\ea~\cite{wang2025:dracogpt}.
Nevertheless, LLMs could expand the diversity of augmented pairs.
Future work might address the above concerns through stronger guarantees or sanity checks of augmented chart pairs.

\bpstart{Editing Knowledge Bases}
While we looked at methods for updating weight terms in a knowledge base, future research should explore additional aspects, including synthesizing novel features and incorporating effect size gaps in empirical studies.
The original data has a few chart pairs without any feature differences.
Furthermore, features may have interaction effects when combined together. 
Future research could explore a reliable approach to extend a knowledge base, not just with updated weights, but with entirely new features. 
Next, across chart pairs, the constituent charts may differ tremendously in terms of effectiveness, but this can be obscured by the use of binary (preferred, not-preferred) labels.
Examining the underlying data from Kim~\ea~\cite{kim2018:encodings}, for example, the effectiveness gap between \textit{x} and color channels is larger than that between color and size. 
Future work could look at how to incorporate more nuanced contrasts within design pairs. 

\bpstart{Beyond Draco}
We demonstrated our augmentation methods via Draco, a knowledge base that encodes design principles as logical rules. 
We discuss implications to different knowledge encoding methods and procedures. 
First, knowledge graph methods (\eg~KG4Vis~\cite{li2022:kg4vis}, AdaVis~\cite{zhang2024:adavis}, GraphScape~\cite{kim2017:graphscape}) could adopt our augmentation methods.
Their entity relations are partite and weighted at a high-level, so we can re-express them as logical expressions and associated weights.
Given relations appearing insufficiently or rarely in a corpus, for example, our feature augmentation could support generating designs with or without those relations.
Then, one could further apply primitive augmentation to those relations.
In addition, our augmentation techniques and negative sampling procedures~\cite{sun2018:rotate} can complement each other to fill in features omitted from initial training data.
Feature augmentation further provides design examples for omitted features.

Next, visualization design recommenders aim to learn a certain design space from charts collected from repositories (\eg~VizML~\cite{hu19:vizml}, MultiVision~\cite{wu2022:multivision}, DMiner~\cite{lin2024:dminer}) or enumerated in a principled way (\eg~GraphScape~\cite{kim2017:graphscape}, LQ$^2$~\cite{wu21:lq2}, Kim~\ea~\cite{kim2021:insights}).
These approaches form a high-level tradeoff: a large pool of real world designs can enhance diversity but may not \textit{guarantee} whether they can fully cover a knowledge base's feature space.
Future work could adapt our approaches in order to collect balanced training data for a more generalizable recommender.
For example, if the goal is to assess a loss function for a complex model, our feature augmentation technique can support this by enumerating pairs with relevant contrasts or guide additional data collection from external sources. 
Alternatively, if the collected designs exhibit low frequencies for certain design characteristics, one could apply our primitive augmentation technique to generate cases by altering the less covered designs with transformations that commonly appear in sufficiently covered designs.

\section*{Supplementary Material}
The supplementary material (\url{https://osf.io/fqpdh/?view_only=4b618c15bebc443db8414c268a934ef7}) includes (1) the code base for data processing, augmentation, auto-labeling, and feature training; and (2) the original and augmented design pairs for the experiment.

%% if specified like this the section will be omitted in review mode
% \acknowledgments{%
% 	The authors wish to thank A, B, and C.
%   This work was supported in part by a grant from XYZ (\# 12345-67890).%
% }

\bibliographystyle{abbrv-doi-hyperref}

\bibliography{references}

\end{document}